\documentclass[reprint,showpacs,preprintnumbers,amsmath,amssymb,twocolumn,superscriptaddress]{revtex4}

	\usepackage{graphicx,dcolumn,bm}
\begin{document}
\title{Rheology of Attractive Emulsions}

\author{Sujit S. Datta}
 \affiliation{Department of Physics, Harvard University, Cambridge MA 02138, USA}
\author{Dustin D. Gerrard}
\affiliation{Department of Physics, Harvard University, Cambridge MA 02138, USA}
\affiliation{Department of Mechanical Engineering, Brigham Young University, Provo UT 84602, USA}
\author{Travers S. Rhodes}
\affiliation{Department of Physics, Harvard University, Cambridge MA 02138, USA}
\author{Thomas G. Mason}
\affiliation{Departments of Chemistry and Physics, and California NanoSystems Institute, University of California, Los Angeles CA 90095, USA}
\author{David A. Weitz}
\email{weitz@seas.harvard.edu}
\affiliation{Department of Physics, Harvard University, Cambridge MA 02138, USA}

\date{\today}

\begin{abstract}
We show how attractive interactions dramatically influence emulsion rheology. Unlike the repulsive case, attractive emulsions below random close packing, $\phi_{RCP}$, can form soft gel-like elastic solids. However, above $\phi_{RCP}$, attractive and repulsive emulsions have similar elasticities. Such compressed attractive emulsions undergo an additional shear-driven relaxation process during yielding. Our results suggest that attractive emulsions begin to yield at weak points through the breakage of bonds, and, above $\phi_{RCP}$, also undergo droplet configurational rearrangements. 
\end{abstract}
\maketitle

\section{Introduction}
Emulsions are suspensions of droplets of one immiscible fluid in another. They are widely used in technological applications requiring the transport and flow of the dispersed fluid; these include oil recovery, food products, pharmacology, coatings, and cosmetics. The droplets are typically stabilized by a surfactant adsorbed on their interfaces; this ensures that there is a short-range repulsive interaction between them, which prevents their coalescence. Such a repulsive emulsion becomes a disordered elastic solid as it is compressed: this behavior can be characterized by the complex shear modulus, $G^{*}(\omega)=G'(\omega)+iG''(\omega)$, where $G'$ is the storage modulus, $G''$ is the loss modulus, and $\omega$ is the angular frequency. For droplet volume fractions, $\phi$, approaching random close packing of spheres, $\phi_{RCP}\approx0.64$, from below, a repulsive emulsion exhibits a weak elasticity that arises from thermal fluctuations \cite{mason1}. By contrast, if $\phi$ is increased above $\phi_{RCP}$, the droplets are forced to deform; as a result, the elasticity is determined by the Laplace pressure scale of the droplets, $\sigma/a$, where $\sigma$ is the interfacial tension between the dispersed and continuous phases and $a$ is the average droplet size \cite{mason1}. Even though it is an elastic solid, such an emulsion can nevertheless be made to flow quite easily: the imposition of sufficiently large shear causes it to yield. This behavior can be elucidated using oscillatory measurements of $G'$ and $G''$ performed at constant $\omega$ and varying maximum strain amplitude $\gamma$. The viscoelastic moduli of a repulsive emulsion above $\phi_{RCP}$ become strain-dependent for sufficiently large $\gamma$ as the emulsion yields. For increasing $\gamma$, the nonlinear $G''$ exhibits a single, well-defined peak at a strain $\gamma^{*}_{r}$ before falling as $\gamma^{-\nu_{r}''}$; $G'$ concomitantly decreases as $\gamma^{-\nu_{r}'}$ with $\nu_{r}''\approx\nu_{r}'/2$ \cite{mason1, mason3, srfs}. The pronounced peak in $G''$ is a characteristic feature of soft glassy materials; it is a direct consequence of a structural relaxation process and thus provides an effective way to characterize yielding \cite{srfs}. This approach is particularly useful for repulsive emulsions above $\phi_{RCP}$, for which the peak in $G''$ reflects the irreversible rearrangements of the densely-packed droplets \cite{mason3, hebraud}.

Another widely-encountered class of emulsions is characterized by droplets with additional attractive interactions between them. In stark contrast to the repulsive case, such an attractive emulsion can form an elastic solid even for $\phi$ well below $\phi_{RCP}$ \cite{encyclopedia}; the bonds between droplets result in a connected network of aggregates that can support a shear stress \cite{encyclopedia, poulin1, bibette2, vanaken}. As a result, an attractive emulsion must exhibit different flow and yielding behavior \cite{colin}. However, despite its broad industrial applications, exactly how this behavior occurs is unknown. Thus, measurements of the characteristic $\gamma$-dependent yielding of attractive emulsions are essential to elucidate how emulsion rheology depends on interdroplet interactions. 

In this Article, we explore the rheology of attractive emulsions using oscillatory measurements over a range of $\phi$. For increasing $\gamma$, $G''(\gamma)$ of attractive emulsions below $\phi_{RCP}$ exhibits a single peak at a strain $\gamma_{1}^{*}\ll\gamma_{r}^{*}$ that increases with $\phi$. By contrast, $G''(\gamma)$ of attractive emulsions above $\phi_{RCP}$ exhibits two peaks at $\gamma^{*}_{1}$ and $\gamma_{2}^{*}\approx\gamma_{r}^{*}$, unlike the repulsive case; these reflect two distinct structural relaxation processes. The time scales of both of these processes vary with shear rate as $\dot{\gamma}^{-\nu}$ with $\nu\approx0.8-1$. Our results provide insight into the elasticity and yielding of attractive emulsions and highlight the sensitivity of emulsion rheology to attractive interactions.

\section{Experimental Details}
We use emulsions comprised of silicone oil droplets dispersed in formamide, a solvent with negligible evaporation; the droplets are sterically stabilized by Pluronic P105, a non-ionic amphiphilic copolymer \cite{copolymersbook}. The mechanical measurements are performed at $T=23^\circ$C on strain- or stress-controlled rheometers (TA ARES G2 or Anton Paar Physica MCR 501, respectively) using a parallel-plate geometry. The sample environment is controlled using a solvent trap. The plates are roughened to eliminate wall-slip, and we verify that measurements are independent of the gap size; furthermore, the results are similar to those obtained using a cone-and-plate geometry, indicating that they are not significantly influenced by the non-uniform strain field characteristic of a parallel-plate geometry. We pre-shear the samples prior to each measurement by imposing a 50s$^{-1}$ constant shear-rate flow for 30s, followed by oscillatory shear of $\gamma$ decreasing from 300\% to 0.01\% over a period of 230s. We do not observe sample creaming over the measurement duration. 

At the temperature and concentrations used here, P105 forms freely-dispersed globular micelles in formamide with radius $a_{m}=6.5$nm, aggregation number $\nu_{m}=31$ and critical micelle concentration $c^{*}=20.2$mM \cite{alexandridis1}. These micelles induce attractive depletion interactions between emulsion droplets \cite{bibette1}, whose magnitude at interdroplet contact $U$ we calculate using the Vrij model \cite{vrij}: $U/k_{B}T=4\pi a a_{m}^{2}N_{A}(c-c^{*})/\nu_{m}$, where $a$ is the average droplet radius, $c$ is the P105 concentration, $k_{B}$ is Boltzmann's constant, $T$ is temperature, and $N_{A}$ is Avogadro's number. These interactions preserve a lubricating layer of formamide between the droplets, and hence are pairwise centrosymmetric and do not resist bending. Because ions do not appreciably self-dissociate in formamide, and because we use a non-ionic surfactant, we do not expect electrostatics to play a significant role \cite{stubenrauch}. 

\begin{figure}
\begin{center}
\includegraphics[width=3.35in]{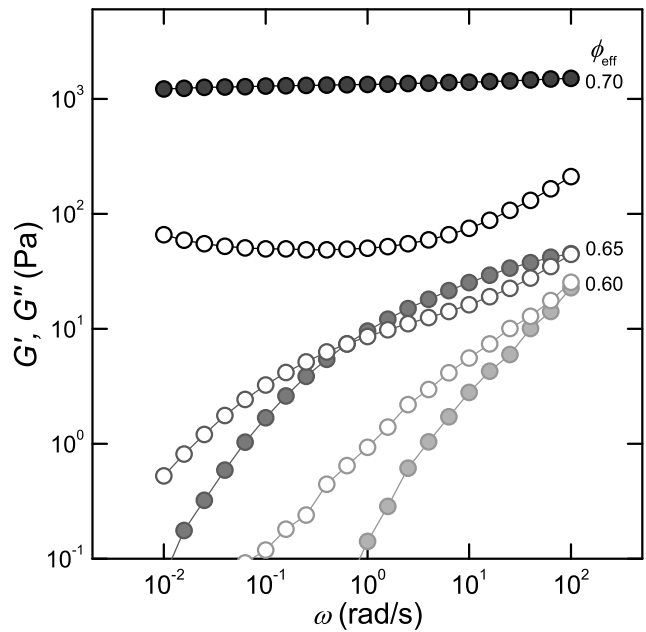}
\caption{{\bf Repulsive emulsions become fluid-like below random close packing.} Linear storage and loss moduli, $G'(\omega)$ and $G''(\omega)$, of repulsive emulsions with $U=0$, $a=128$nm, and $\phi_{eff}\approx$ 0.70, 0.65, and 0.60 (progressively lighter colors). } 
\end{center}
\end{figure}

\begin{figure}
\begin{center}
\includegraphics[width=3.35in]{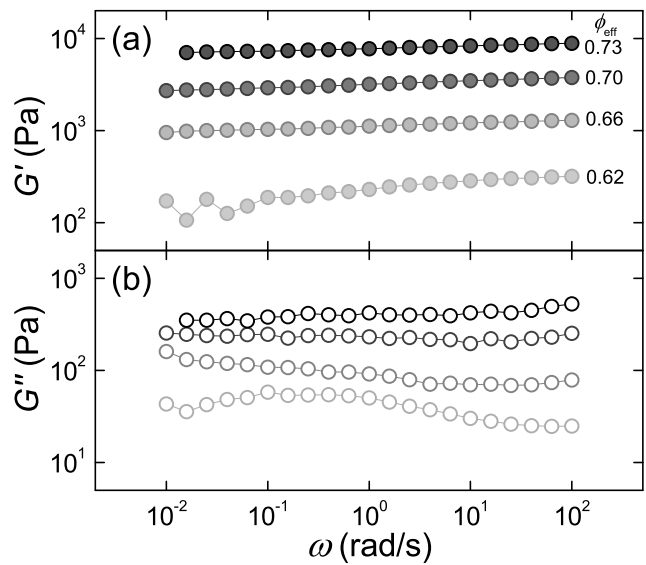}
\caption{{\bf Attractive emulsions are elastic both above and below random close packing.} Linear storage and loss moduli, (a) $G'(\omega)$ and (b) $G''(\omega)$, of attractive emulsions with $U\approx9k_{B}T$, $a=100$nm, and $\phi_{eff}\approx$ 0.73, 0.70, 0.66, and 0.62 (progressively lighter colors). For clarity, $G'$ and $G''$ data are multiplied by a factor of 0.5, 0.2, or 0.07 for  $\phi_{eff}=0.70, 0.66,$ or 0.62, respectively.} 
\end{center}
\end{figure}

To investigate the influence of attraction on the rheology of emulsions, we prepare repulsive emulsions with $U=0$ ($c=7$mM$~<c^{*}$) and attractive emulsions with $U=7-9k_{B}T$ ($c=26.8-28.7$mM) over a range of $\phi$ using high-shear rate homogenization or sonication with a probe tip. We densify the emulsions using centrifugation and verify that they are stable against coalescence or Ostwald ripening \cite{pine}. We measure $\sigma=6-9$mN/m using a du No\"uy ring. The oil volume fraction $\phi$ is determined by measuring the masses of all components making up a sample. The droplets have $a=100, 106,$ or $128$nm and relatively low polydispersity $\sim30-35\%$ as measured using dynamic light scattering. We also account for the thickness $t$ of the surfactant layer adsorbed at the droplet surfaces to obtain the effective volume fraction $\phi_{eff}\approx\phi(1+t/(a-t))^{3}$. We estimate $t\approx3.1$nm based on neutron scattering data \cite{alexandridis1,copolymersbook} and further verify this by fitting viscosity measurements of dilute repulsive samples to simulation data appropriate for emulsions \cite{mason3, ladd}. This estimate for $t$ is likely to be unchanged even at the highest $\phi$ studied: the pressure required to compress the surfactant layer $\sim k_{B}T/s^{3}$, where $\pi s^{2}$ is the interfacial area per polymer molecule, is approximately two orders of magnitude larger than the maximum osmotic pressure exerted on the droplets $\sim0.1\sigma/a$ \cite{degennes1,mason1}.

To test the generality of our results, we also study silicone oil-in-water emulsions stabilized by an ionic surfactant, SDS. All of the data presented in this paper are for the oil-in-formamide emulsions stabilized by P105, with the exception of Figs. 7 and 8, which present data for the oil-in-water emulsions stabilized by SDS. These are well-defined and stable model systems characterized by very weak ($<1k_{B}T$) or strong ($>1k_{B}T$) attractive interdroplet interactions using an SDS concentration only slightly greater than or significantly greater than the critical micelle concentration ($c^{*}\approx8$mM), respectively \cite{bibette1,mason1,mason2}. We prepare the emulsions using depletion fractionation; the droplets have $a=250$nm and have low polydispersity $\sim10\%$. To investigate the influence of attraction on the rheology of emulsions, we study repulsive emulsions with $U<1k_{B}T$ ($c=10$mM) and attractive emulsions with $U\approx21k_{B}T$ ($c=200$mM). The effective volume fraction $\phi_{eff}$ is defined to incorporate the effects of the thin liquid film between adjacent droplets due to their electrostatic repulsion. We study the emulsions using oscillatory rheology on a strain-controlled rheometer using a roughened cone-and-plate geometry with a vapor trap. Further details and the rheology data for the SDS-stabilized repulsive emulsions are presented in \cite{mason1,mason2}. 

\section{Results and Discussion}

\begin{figure}
\begin{center}
\includegraphics[width=3.46in]{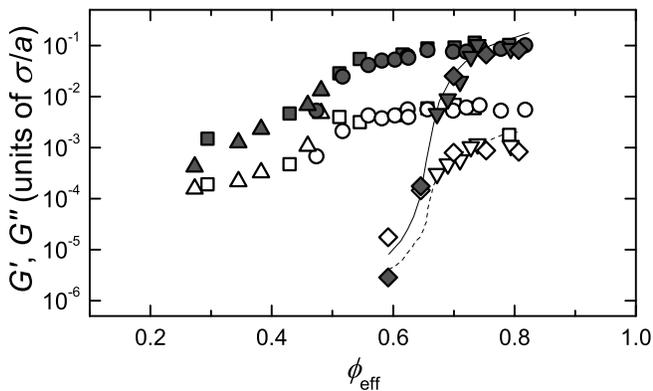}
\caption{{\bf Elasticities of attractive and repulsive emulsions differ below random close packing but are similar above it.} Linear storage and loss moduli, $G'_{p}$ (solid points) and $G''$ (open points), measured at $\omega=1$ rad/s, normalized by the Laplace pressure $\sigma/a$, for both attractive and repulsive emulsions of varying $\phi_{eff}$. Attractive emulsions have $a=128$nm, $U\approx9k_{B}T$ (circles), $a=106$nm, $U\approx7k_{B}T$ (upward triangles), and $a=100$nm, $U\approx9k_{B}T$ (squares); repulsive emulsions have $a=128$nm (diamonds) and $a=106$nm (downward triangles). The interfacial tension $\sigma=6-9$mN/m for the range of surfactant concentrations used. Previous measurements of $G'_{p}$ (solid line) and $G''$ (dashed line) for monodisperse repulsive emulsions \cite{mason1}, horizontally shifted to account for polydispersity, agree with our data. We find $\phi_{RCP}\approx0.68-0.72$, characteristic of $\phi_{RCP}$ for spheres of comparable polydispersity \cite{farrgroot}.}
\end{center}
\end{figure}

\begin{figure}
\begin{center}
\includegraphics[width=3.35in]{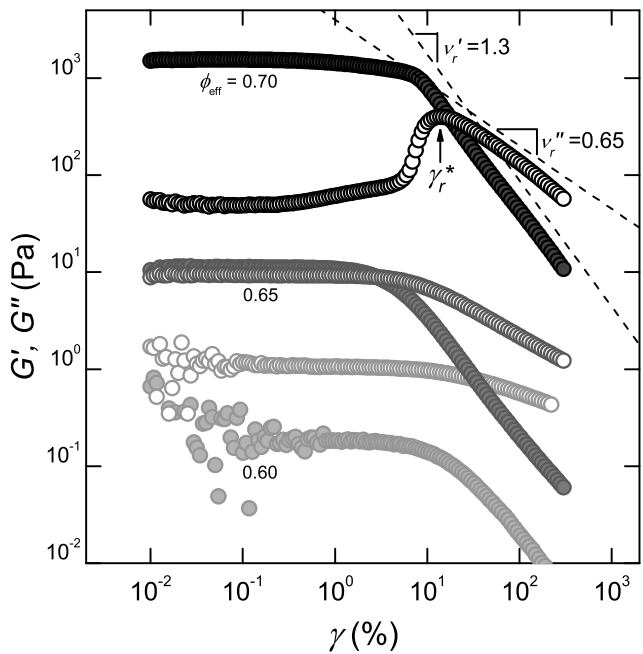}
\caption{{\bf Repulsive emulsions above random close packing exhibit one peak in $G''(\gamma)$ at $\gamma_{r}^{*}$ while those below random close packing do not exhibit a peak in $G''(\gamma)$.} Viscoelastic moduli $G'(\gamma)$ and $G''(\gamma)$ measured at $\omega=1$ rad/s of repulsive emulsions with $U=0$, $a=128$nm, and $\phi_{eff}\approx$ 0.70, 0.65, and 0.60 (progressively lighter colors). }
\end{center}
\end{figure}

\begin{figure}
\begin{center}
\includegraphics[width=3.35in]{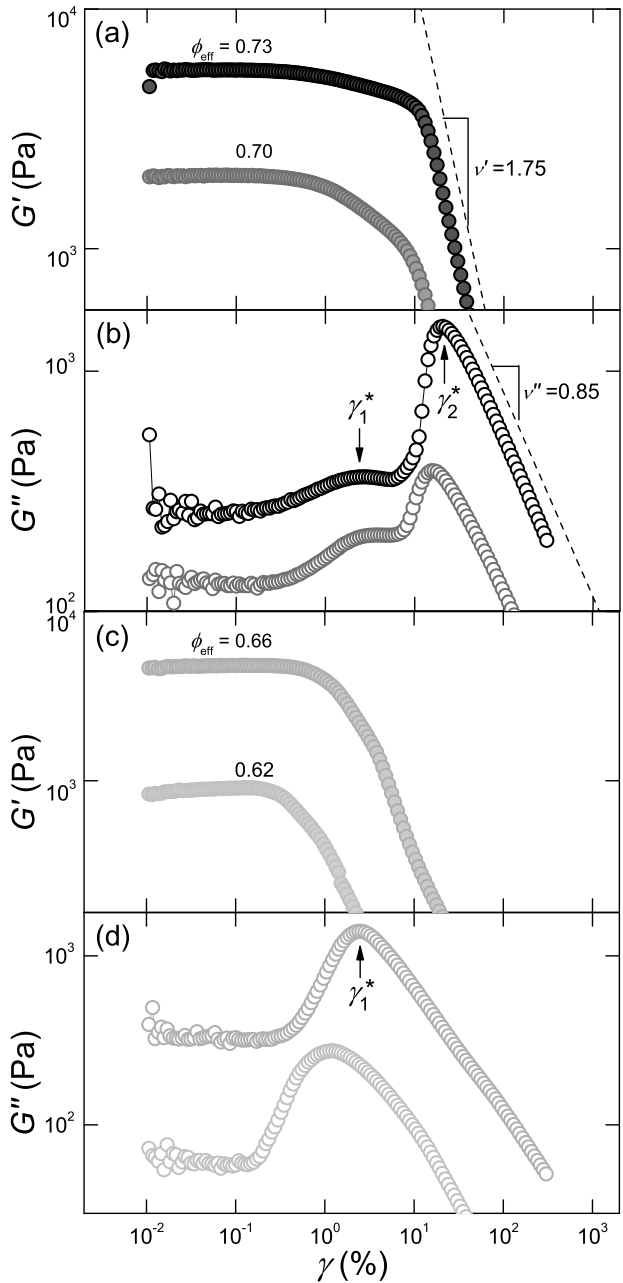}
\caption{{\bf Attractive emulsions above random close packing exhibit two peaks in $G''(\gamma)$ at $\gamma_{1}^{*}$ and $\gamma_{2}^{*}$ while those below random close packing exhibit one peak at $\gamma_{1}^{*}$.} Viscoelastic moduli $G'(\gamma)$ and $G''(\gamma)$ measured at $\omega=1$ rad/s of attractive emulsions with $U\approx9k_{B}T$ and $a=100$nm (a-b) above $\phi_{RCP}$ ($\phi_{eff}\approx 0.73$ and 0.70, progressively lighter colors) and (c-d) below $\phi_{RCP}$ ($\phi_{eff}\approx 0.66$ and 0.62, progressively lighter colors). For clarity, $G'$ and $G''$ data are multiplied by a factor of 0.7, 0.3, 0.8, or 0.2 for each $\phi_{eff}$ going from top to bottom. Straight lines indicate $G',G''\sim\gamma^{-\nu',-\nu''}$ for large $\gamma$. }
\end{center}
\end{figure}

The elastic behavior of emulsions is characterized by a $\omega$-independent regime of $G'(\omega)$; two examples, a repulsive emulsion with $\phi_{eff}=0.70$ and an attractive emulsion with $\phi_{eff}=0.73$, are shown in the top curves of Figs. 1 and 2, respectively. Repulsive emulsions are solid-like when packed above $\phi_{RCP}$ but become fluid-like as $\phi_{eff}$ is decreased below $\phi_{RCP}$ [Fig. 1] \cite{mason1}. By contrast, attractive emulsions are solid-like over a wider range of $\phi_{eff}$ [Fig. 2]. To summarize this behavior, we plot the plateau modulus $G'_{p}$, measured at $\omega=1$ rad/s, as a function of $\phi_{eff}$. Above $\phi_{RCP}$, $G'_{p}$ has the same magnitude for both attractive and repulsive emulsions as shown in Fig. 3. This indicates that the elasticity in both cases is dominated by the repulsive forces deforming the droplets. However, $G''$ is an order of magnitude larger for attractive emulsions, as compared to the repulsive case [Fig. 3]. For repulsive emulsions, $G'_{p}$ drops precipitously as $\phi_{eff}$ is decreased below $\phi_{RCP}$, indicating that the droplets are no longer compressed \cite{mason1}; by contrast, the elasticity of attractive emulsions persists far below $\phi_{RCP}$, as shown by the upper symbols in Fig. 3.

To elucidate the microscopic mechanisms for emulsion flow, we investigate the yielding of emulsions with different $\phi_{eff}$ by measuring the $\gamma$-dependence of $G'$ and $G''$ at $\omega=1$ rad/s. Repulsive emulsions above $\phi_{RCP}$ begin to yield at $\gamma^{*}_{r}\sim10\%$ and $G''$ exhibits a single, well-defined peak at this strain, as indicated by the arrow in Fig. 4 \cite{mason1, mason3, srfs}. Attractive emulsions above $\phi_{RCP}$ begin to yield at much smaller strain. Unexpectedly, $G''$ exhibits two well-defined peaks, a first at  $\gamma^{*}_{1}\sim1\%\ll\gamma_{r}^{*}$ and a second at $\gamma_{2}^{*}\approx\gamma_{r}^{*}$, before falling as $\gamma^{-\nu''}$, as indicated by the arrows in Fig. 5(b). Correspondingly, $G'$ decreases weakly for $\gamma>0.5\%$ before falling as $\gamma^{-\nu'}$ for $\gamma>10\%$, with $\nu''\approx\nu'/2$ [Fig. 5(a)]. By contrast, $G''$ of attractive emulsions below $\phi_{RCP}$ only exhibits a single peak at $\gamma^{*}_{1}$, as shown in Fig. 5(d), and $G'$ decreases smoothly for $\gamma>0.01-1\%$ [Fig. 5(c)].

\begin{figure}
\begin{center}
\includegraphics[width=3.46in]{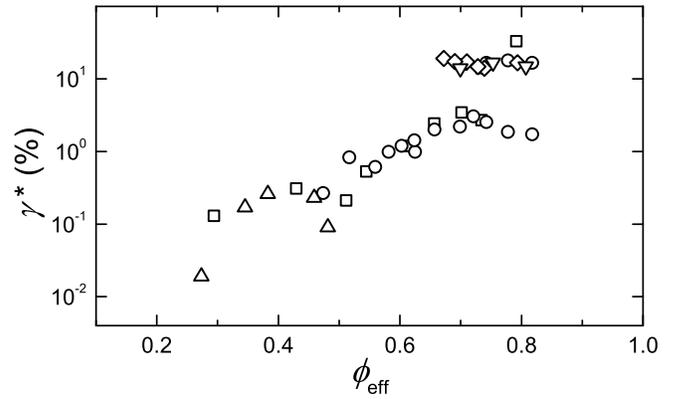}
\caption{{\bf Position of peaks in $G''(\gamma)$ for varying $\phi_{eff}$.} Lower symbols represent $\gamma_{1}^{*}$, upper symbols represent either $\gamma_{2}^{*}$ or $\gamma_{r}^{*}$. Attractive emulsions have $a=128$nm, $U\approx9k_{B}T$ (circles), $a=106$nm, $U\approx7k_{B}T$ (upward triangles), and $a=100$nm, $U\approx9k_{B}T$ (squares); repulsive emulsions have $a=128$nm (diamonds) and $a=106$nm (downward triangles).}
\end{center}
\end{figure}

To understand these results, we examine the $\phi_{eff}$-dependence of $\gamma_{1}^{*}$. Below $\phi_{RCP}$, $\gamma_{1}^{*}$ decreases with decreasing $\phi_{eff}$, as shown by the lower symbols in Fig. 6 \cite{stress}; interestingly, this is similar to behavior predicted for particulate colloidal gels connected by ``weak links" \cite{aksay}. The elasticity of these emulsions results from the attractive interdroplet bonds; these induce the formation of a stress-bearing connected network comprised of compact droplet aggregates \cite{poulin1, bibette2, vanaken}. We expect the elasticity to be dominated by the weakest bonds in the network, and hence yielding begins when these are broken. In this picture, a macroscopic deformation $\Delta L$ deforms such a bond by $\Delta L/(L/\zeta)$, where $L$ is the system size and $\zeta$ is the characteristic distance between the weakest bonds. The force on such a bond is thus $k_{a}\Delta L/(L/\zeta)=k_{a}\zeta\gamma$, where $k_{a}$ is the characteristic bond stiffness. Assuming that the bond breaks at a fixed critical force $F^{*}=k_{a}\zeta\gamma_{1}^{*}$, the critical strain amplitude $\gamma_{1}^{*}\propto\zeta^{-1}$. Our measurements of $\gamma_{1}^{*}$ thus imply that $\zeta$ increases with decreasing $\phi_{eff}$, suggesting that yielding begins at fewer, sparser weak points as $\phi_{eff}$ is decreased. This may reflect the changing connectivity of the emulsion: the characteristic size $\xi$ of the droplet aggregates comprising the emulsion also increases with decreasing $\phi_{eff}$ \cite{poulin1,vanaken, confirm}. If the bonds between droplet aggregates are the weakest in the emulsion \cite{delgado, laurati}, we would expect $\zeta\approx\xi$ to increase with decreasing $\phi_{eff}$, consistent with our results.

\begin{figure}
\begin{center}
\includegraphics[width=3.20in]{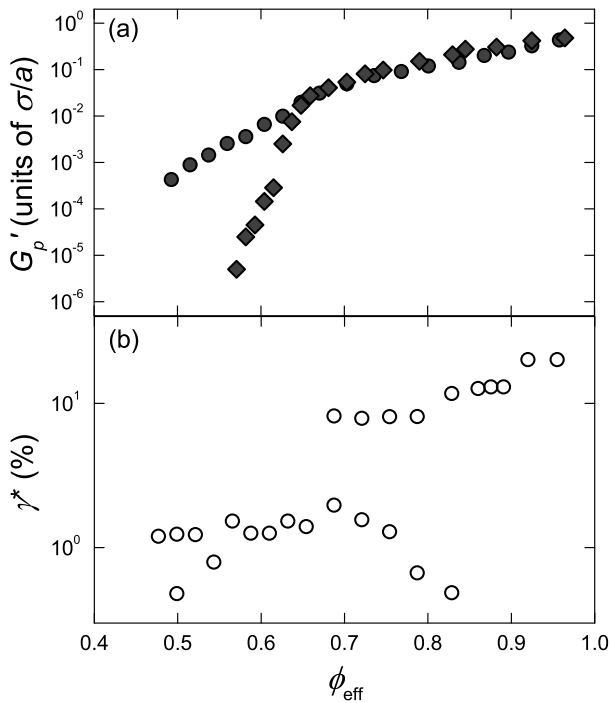}
\caption{{\bf Volume fraction-dependent behavior is similar for oil-in-water emulsions stabilized by SDS.} (a) Linear plateau storage modulus, $G'_{p}$ \cite{mason2}, normalized by the Laplace pressure $\sigma/a$ and (b) position of peaks in $G''(\gamma)$, for varying $\phi_{eff}$. Emulsions have $a=250$nm; attractive emulsions have $U\approx21k_{B}T$ (circles) while repulsive emulsions have $U<1k_{B}T$ (diamonds). The interfacial tension $\sigma=9.8$mN/m for the range of surfactant concentrations used. Upper symbols in (b) represent $\gamma_{2}^{*}$ while lower symbols represent $\gamma_{1}^{*}$.} 
\end{center}
\end{figure}

\begin{figure}
\begin{center}
\includegraphics[width=3.45in]{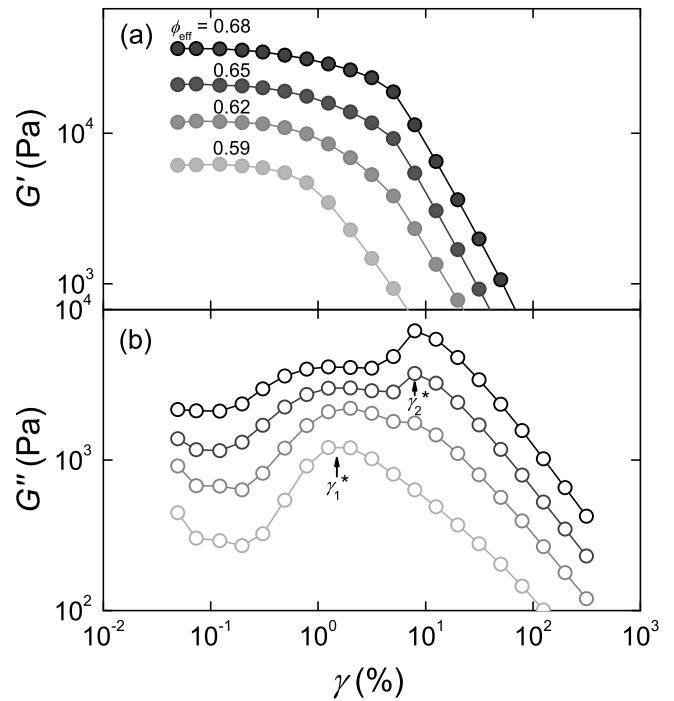}
\caption{{\bf Strain-dependent behavior is similar for oil-in-water emulsions stabilized by SDS.} Viscoelastic moduli (a) $G'(\gamma)$ and (b) $G''(\gamma)$ of attractive emulsions stabilized by SDS with $U\approx21k_{B}T$ and $a=250$nm for $\phi_{eff}\approx 0.68$, 0.65, 0.62, and 0.59 (progressively lighter colors). Emulsions above $\phi_{RCP}$ (top two curves) show peaks in $G''$ at $\gamma_{1}^{*}$ and $\gamma_{2}^{*}$ while those below $\phi_{RCP}$ (lower two curves) show a single peak at $\gamma_{1}^{*}$. For clarity, $G'$ and $G''$ data are multiplied by a factor of 0.7, 0.6, or 0.5 for  $\phi_{eff}=0.65, 0.62,$ or 0.59, respectively.} 
\end{center}
\end{figure}

\begin{figure}
\begin{center}
\includegraphics[width=3.0in]{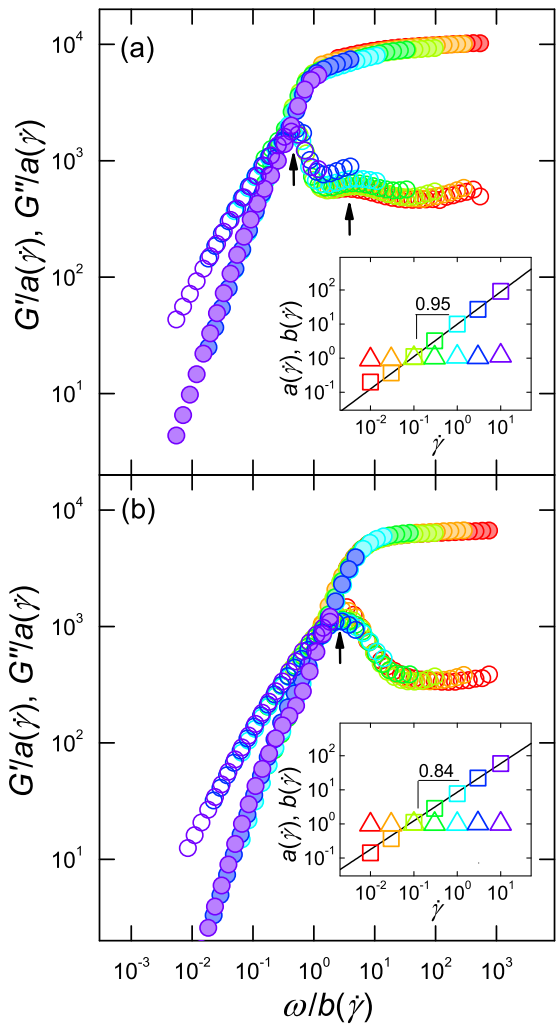}
\caption{(Color Online) {\bf Yielding of attractive emulsions is a shear-driven process.} Constant $\dot{\gamma}$ frequency sweep measurements of viscoelastic moduli $G'(\omega)$ and $G''(\omega)$ for two samples with $a=100$nm and $U\approx9k_{B}T$ above $\phi_{RCP}$ (a, $\phi_{eff}\approx0.73$) and below $\phi_{RCP}$ (b, $\phi_{eff}\approx0.66$), shifted onto a single master curve by normalizing by the shear-rate-dependent shift factors $a(\dot{\gamma})$ and $b(\dot{\gamma})$. Inset: corresponding amplitude and frequency shift factors $a(\dot{\gamma})$ (triangles) and $b(\dot{\gamma})$ versus $\dot{\gamma}$ (squares). }
\end{center}
\end{figure}

For increasing $\gamma$, attractive emulsions both below and above $\phi_{RCP}$ begin to yield and exhibit a peak in $G''$ at $\gamma\sim\gamma^{*}_{1}$; in contrast, those above $\phi_{RCP}$ exhibit an additional peak in $G''$ at $\gamma\sim\gamma^{*}_{2}$ [Fig. 5]. This observation implies the existence of an additional process by which the structure of these emulsions must relax before they can flow. This is likely due to the compression of the droplets: above $\phi_{RCP}$, attractive emulsions require larger strain to fully yield through the irreversible rearrangements of densely-packed droplets, just as in the repulsive case \cite{mason3,hebraud}. This hypothesis is supported by our observation that $\gamma_{2}^{*}\approx\gamma_{r}^{*}$. The height of the first yielding peak in $G''(\gamma)$ decreases as $\phi_{eff}$ increases above $\phi_{RCP}$, as shown in Fig. 5(b), reflecting the increasing relative importance of these repulsive interactions as the droplets are increasingly compressed \cite{procaccia}.  

To test the generality of these results, we perform similar measurements on an oil-in-water emulsion stabilized by a different surfactant. The droplets are electrostatically stabilized by SDS, an ionic surfactant that also forms freely-dispersed globular micelles, and have low polydispersity ($\sim10\%$) \cite{mason1,mason2}. Similar to the case of P105-stabilized emulsions, $G'_{p}$ measured for the SDS-stabilized emulsions has the same magnitude for both attractive and repulsive emulsions packed above $\phi_{RCP}\approx0.64$ [Fig. 7(a)]; this indicates that the elasticity is dominated by the repulsive forces deforming the droplets. For the repulsive emulsions, $G'_{p}$ drops precipitously as $\phi_{eff}$ is decreased below $\phi_{RCP}$; by contrast, the elasticity of the attractive emulsions persists far below $\phi_{RCP}$ [Fig. 7(a)]. Moreover, $G''(\gamma)$ of SDS-stabilized attractive emulsions above $\phi_{RCP}$ exhibits two peaks at strains $\gamma_{1}^{*}\sim1\%\ll\gamma_{r}^{*}$ and $\gamma_{2}^{*}\approx\gamma_{r}^{*}$, as indicated in Fig. 8(b). Correspondingly, $G'$ decreases weakly for $\gamma>0.5\%$ before falling more quickly for $\gamma>10\%$ [Fig. 8(a)]. By contrast, $G''(\gamma)$ of the attractive emulsions below $\phi_{RCP}$ exhibits a single peak at a strain $\gamma_{1}^{*}$, and $G'$ decreases smoothly for $\gamma>1\%$  [Fig. 8]. These data are similar to those obtained for P105-stabilized emulsions; this confirms that our results are more general. 

The frequency-dependent mechanical response of emulsions directly reflects the time scales of their structural relaxation. For repulsive emulsions above $\phi_{RCP}$, $G''(\omega)$ exhibits a shallow minimum at $\omega\sim0.1-10$ rad/s resulting from the combination of viscous loss at high $\omega$ and the configurational rearrangements of the individual droplets at low $\omega$ [Fig. 1] \cite{mason1, srfs}. In contrast, for attractive emulsions, $G''(\omega)$ exhibits a minimum at $\omega_{min}\geq10$ rad/s, as shown in Fig. 2. This further suggests that attractive interactions alter the structural relaxation of emulsions. 

The structural relaxation process through which attractive emulsions yield occurs at frequencies much lower than is accessible in a linear rheological measurement, typical of soft glassy materials \cite{srfs}. We circumvent this problem by performing $\omega$-dependent measurements holding the strain rate amplitude $\dot{\gamma}=\gamma\omega$ constant for different values of $\dot{\gamma}$. A relaxation process of time scale $\tau$ gives rise to a peak in $G''(\omega)$ at a frequency proportional to $\tau^{-1}$ \cite{srfs}. We observe one peak in $G''(\omega)$ for attractive emulsions with $\phi_{eff}<\phi_{RCP}$; strikingly, we observe two peaks in $G''(\omega)$ for those with $\phi_{eff}>\phi_{RCP}$, as indicated in Fig. 9, in stark contrast to the repulsive case \cite{srfs}. This provides further confirmation that attractive emulsions above $\phi_{RCP}$ undergo an additional relaxation process during yielding. 

The data measured at different $\dot{\gamma}$ can be scaled onto a master curve by rescaling the moduli and frequency, as shown in Fig. 9. The frequency scaling factor $b(\dot{\gamma})\sim\tau^{-1}$ \cite{srfs}. We explore its dependence on $\dot{\gamma}$ to directly probe how the time scales of the structural relaxation processes of attractive emulsions depend on shear rate.  We find $b(\dot{\gamma})\sim\tau^{-1}\sim{\dot{\gamma}}^{\nu}$ with $\nu\approx0.8-1$ for both relaxation processes in attractive emulsions having $\phi_{eff}$ both below and above $\phi_{RCP}$; two representative examples are shown in Fig. 9 (inset). This shear-driven behavior is similar to the yielding of colloidal gels \cite{sprakel} and to the yielding of repulsive emulsions above $\phi_{RCP}$ \cite{srfs}.
  
\section{Conclusion}
Our data suggest a simple physical picture of emulsion flow. The rheology of emulsions with slippery bonds of magnitude $U<1k_{B}T$ is similar to that of repulsive emulsions; in stark contrast, emulsions with $U>7k_{B}T$ show a dramatically enhanced elasticity below $\phi_{RCP}$. Moreover, their nonlinear rheology is markedly different from the repulsive case. These attractive emulsions begin to yield under sufficient shear through the breakage of interdroplet bonds at weak points in the emulsion \cite{colin}. Attractive emulsions above $\phi_{RCP}$ also undergo shear-induced configurational rearrangements of the densely-packed droplets, similar to the repulsive case, at larger strain. This is reminiscent of the two-step yielding of attractive colloidal glasses \cite{pham1}. How this behavior changes with intermediate values of $U$ remains to be explored.

Emulsions are often used to model many other diverse physical systems and are crucial in various technological processes. We find that their linear and nonlinear rheology depend sensitively on the interactions between the droplets. Our results may thus be a useful input to such models and could help guide processes designed to control emulsion elasticity and flow behavior.

\begin{acknowledgements}
It is a pleasure to acknowledge E. Del Gado for useful discussions and a critical reading of the manuscript; C. O. Osuji, V. Trappe, and J. N. Wilking for useful discussions; and the anonymous referees for useful suggestions. This work was supported by the NSF (DMR-1006546), the Harvard MRSEC (DMR-0820484), the Harvard MRSEC REU, and a fellowship from ConocoPhillips.
\end{acknowledgements}


\begin{thebibliography}{10}
\bibitem{confirm}
Our optical microscopy measurements (not shown) are consistent with this
  finding.
\end{thebibliography}
	\end{document}